\documentclass[twocolumn,journal]{IEEEtran}
\usepackage[T1]{fontenc}
\usepackage[latin9]{inputenc}
\usepackage{float}
\usepackage{mathrsfs}
\usepackage{amstext}
\usepackage{graphicx}

\usepackage{fancyhdr}
 
\usepackage{esint}
\usepackage[unicode=true,
 bookmarks=true,bookmarksnumbered=true,bookmarksopen=true,bookmarksopenlevel=1,
 breaklinks=false,pdfborder={0 0 0},pdfborderstyle={},backref=false,colorlinks=false]
 {hyperref}
\hypersetup{pdftitle={Your Title},
 pdfauthor={Your Name},
 pdfpagelayout=OneColumn, pdfnewwindow=true, pdfstartview=XYZ, plainpages=false}
\usepackage{breakurl}

\makeatletter
\let\oldforeign@language\foreign@language
\DeclareRobustCommand{\foreign@language}[1]{%
  \lowercase{\oldforeign@language{#1}}}

\usepackage[caption=false,font=footnotesize]{subfig}

\makeatother

\begin{document}

\title{An Accurate Analytic Model for Traveling Wave Tube Dispersion Relation}

\author{\IEEEauthorblockN{Ahmed F. Abdelshafy$^{*}$, Filippo Capolino$^{*}$, and Alexander
Figotin$^{\dagger}$} \\
\IEEEauthorblockA{$^{*}$\textit{Department of Electrical Engineering and Computer Science,
University of California, Irvine, CA 92697 USA} \\
$^{\dagger}$\textit{Department of Mathematics, University of California,
Irvine, CA 92697 USA} \\
 abdelsha@uci.edu and f.capolino@uci.edu, afigotin@uci.edu}}

\markboth{}{Your Name \MakeLowercase{\emph{et al.}}: Your Title}
\maketitle

\thispagestyle{fancy}

\begin{abstract}
We construct an analytical model for the dispersion of the hot modes
in a traveling wave tube (TWT) based on the Lagrangian field theory,
upgrading its constants to be frequency-dependent. The frequency dependence
of the parameters of the TWT slow wave structure (SWS) is recovered
from full-wave simulations by standard software (e.g., CST). We applied
the model to study the hot modes of a helical-based TWT and found
an excellent agreement between the results from our model and those
from particle in cell (PIC) simulations. Our additional studies show
that the proposed approach can be applied to various SWS geometries.
\end{abstract}

\begin{IEEEkeywords}
traveling wave tubes, slow wave structures, dispersion relation, particle
in cell
\end{IEEEkeywords}

\section{Introduction}

Vacuum electron devices (VEDs) are used widely for radar and satellite
communications applications for several decades due to their high
power operational capabilities and reliability\cite{benford_high_2015,gilmour_klystrons_2011}.
VEDs operation is based on the synchronous interaction of an electromagnetic
(EM) wave in a slow wave structure (SWS) with the electron beam \cite{pierce_theory_1947,pierce_travelling-wave_1950}.

We advance here an analytical model of traveling wave tubes (TWTs)
based on the Lagrangian field as in \cite{figotin_multi-transmission-line-beam_2013,figotin2020analytic}
upgrading its constants to be frequency-dependent. The frequency dependence
of the cold circuit parameters, like modal phase velocity and characteristic
wave impedance,  was already included previously in TWTs computational
models, e.g.,  in \cite{wohlbier2002multifrequency,converse2004impulse}.
We refer to an eigenmode as ``hot'' if it is the one associated
with the full-interactive TWT system and as ``cold'' if it is associated
with the SWS (without the presence of the electron beam). The hot
eigenmodes involve both the charge wave and the EM fields and may
have a complex-valued wavenumber, and determining them is the main
subject of our study. The commonly taken initial step in the studies
of TWT eigenmodes is to consider the cold eigenmodes in the SWS in
order to establish conditions providing synchronous interaction between
the charge wave on the electron beam and the EM wave in the SWS. It
is well known that the synchronization occurs when the speed of the
EM wave in the SWS matches the speed of the beam electrons to facilitate
effective energy transfer\cite{pierce_travelling-wave_1950,tsimring_electron_2006,gilmour_klystrons_2011}.
While the studies of the EM eigenmodes in the cold SWS are important
for making good choices when designing the TWT, its actual efficiency
is fully manifested only in hot eigenmodes. The hot eigenmodes, in
particular, carry significant information on electron beam instabilities
such as convective and absolute instability \cite{sturrock_kinematics_1958},
which are crucial for electron beam energy harvesting (i.e., the energies
transfer from the electron beam to the EM wave). The hot mode exponential
growth in space is expressed through the relevant complex-valued wavenumber
with non-zero imaginary part representing the TWT gain.

In most cases, a theoretical model or a computer simulation is used
to model and design TWTs. Pierce's classical small-signal theory has
been widely used for modeling and designing TWTs for about seventy
years \cite{pierce_circuits_1949,pierce_travelling-wave_1950}. Pierce
used the 3-wave theory and described the dispersion relation as a
cubic polynomial (i.e., also know as 3-wave dispersion) \cite{pierce_theory_1947}
that is fully characterized by four dimensionless constants. Many
earlier works are carried out to quantitatively evaluate those constants
(\cite{lau1992review,simon2017evaluation}, and references therein).
Other studies have extended the work by Pierce to theoretically model
TWTs as in \cite{sturrock_kinematics_1958,tamma_extension_2014}.
Recently, a numerical eigenmode solver for hot eigenmodes in TWT systems
was introduced based fully on PIC simulations in \cite{mealy2020traveling},
that is however more complicated than the method proposed here since
in this work we constrain the dispersion to follow the physics predicted
by an analytical model. In spite of being excellent tools for the
initial design of TWTs, theoretical models are often not reliable,
and the actual TWT performance is determined by accurate and time
consuming partricle-in-cell (PIC) simulations. In this research, we
present a mean to narrow the gap between the theoretical predictions
and PIC simulation results.

One of the focuses of our efforts, is the recovering of the information
about hot eigenmodes from a very reduced set of PIC simulations (as
explained later on, a single PIC simulation at only one frequency
is sufficient). The biggest challenge in addressing this subject is
that the extraction of useful information about eigenmodes from PIC
full-wave simulations is not by any means a simple straightforward
problem. We address the problem by a thoughtful selection of special
regimes of TWT operation in which the hot eigenmode features are manifested
in the most pronounced and undisturbed form. The proposed approach
utilizes information of key physical quantities obtained in PIC simulations,
such as EM fields and electrons' energy, and it allows to extract
the spectral information in the form of complex-valued wavenumbers
as well as harmonics of the hot Floquet-Bloch eigenmodes.

\begin{figure*}[!t]
\begin{centering}
\includegraphics[width=0.7\paperwidth]{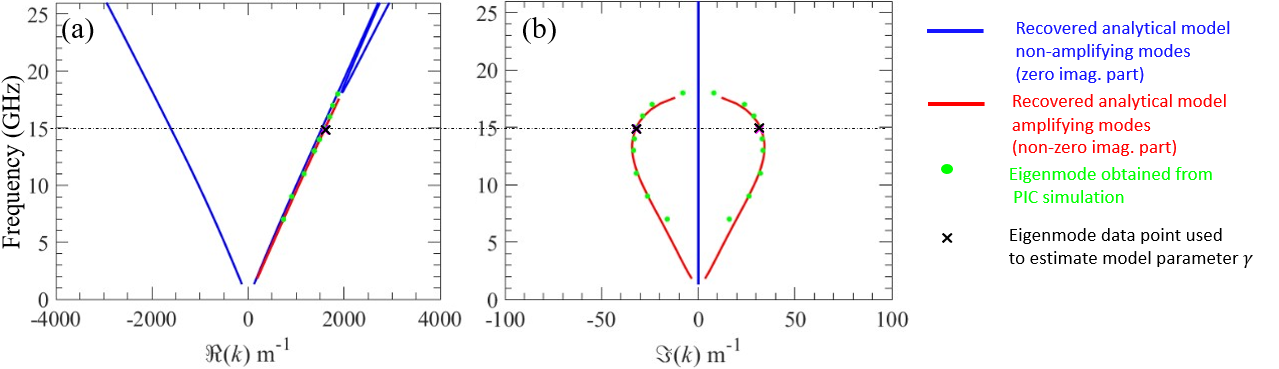}
\par\end{centering}
\caption{\label{fig:After-adjust_dispersion-into} Complex-valued wavenumbers
of the hot eigenmodes in the interactive electron beam-EM mode in
the helical SWS in Fig. \ref{fig:Helical-TWT,-cold sim}(a). (a-b)
Real-imaginary parts of the recovered dispersion by estimating the
adjusted analytical model ($\gamma$) parameter using PIC simulations
data (the black cross data point) and the frequency dependence of
cold wavenumber $w(\omega)$ and $\gamma(\omega)$, directly obtained
from the cold CST eigenmode solver in Fig. \ref{fig:Phase velcoity}.
A good agreement is observed in the real part in (a) as well as the
imaginary part in (b) between the recovered adjusted analytical model
and PIC simulations data (the green dots).}
\end{figure*}

The above mentioned spectral information allows in turn to reconstruct
the analytical model parameters and its frequency dependence. From
the analytical model we obtain the complex-valued ($\omega$-$k$)
dispersion relation of the hot eigenmodes in the TWT (i.e., in the
beam-EM interactive system). This dispersion relation is compared
side by side with data obtained from PIC simulations at various frequencies.
As one can see from Fig. \ref{fig:After-adjust_dispersion-into},
the dispersion relation based on the recovered analytical model is
in excellent agreement with results from three dimensional PIC multi-frequency
simulations for the helical-based TWT shown in Fig. \ref{fig:Helical-TWT,-cold sim}.
The dispersion relation and the frequency dependence of the imaginary
part of the complex-valued wave numbers provide valuable information
on the TWT operation. For instance, the amplification region refers
to the range of the frequencies where $\Im(k)>0$ and the TWT gain
is directly related to $\Im(k)$. The constructed analytical model
with frequency dependence parameters accomplishes our primary goal
to model TWT operations for a wide range of frequencies accurately
as verified by the comparison with PIC simulations. Taking into account
the complexity of TWT operation and its simulation, the simplicity
of the analytical model and its excellent agreement with PIC simulations
is rather remarkable.

The remainder of this article is organized as follows. Section II
presents a brief review of the analytical model used to describe the
system and associated model parameters. Section III presents the proposed
approach to obtain hot eigenmodes of interest in a TWT using a single
PIC simulation (at one frequency). Section IV demonstrates the connection
between the analytical model and actual electron beam device based
on eigenmode and dispersion relations analysis. Conclusions are presented
in Section V.

\section{Mathematical Formulation\label{sec:Analytical-model}}

We briefly review the analytical model of TWTs used in this paper.
An effective mathematical model for a TWT was introduced by Pierce
\cite{pierce_circuits_1949,pierce_travelling-wave_1950}. This model
can be considered as the simplest one that accounts for EM wave amplification
and the electron beam energy conversion into microwave radiation in
the TWT \cite{gilmour_principles_1994,tsimring_electron_2006,gilmour_klystrons_2011}.
The Pierce model, also known as the 4-wave theory of a TWT, is one-dimensional
linear theory in which the SWS is represented by a lossless transmission
line (TL), assumed to be homogeneous, that is, with uniformly distributed
capacitance and inductance \cite{pierce_theory_1947,pierce_travelling-wave_1950,tsimring_electron_2006}.
An approximation to the 4-wave theory is the 3-wave small-signal theory
which laid the foundation for TWT design \cite{pierce_travelling-wave_1950}.

The analytical model used in this paper is a generalization of the
Pierce theory, based on the Lagrangian field framework presented in
Refs. \cite{figotin_multi-transmission-line-beam_2013,figotin2020analytic},
that, in particular, takes into account the space charge effects.
The Lagrangian field theory in \cite{figotin_multi-transmission-line-beam_2013,figotin2020analytic},
allows also to model more complex SWSs than the simple one represented
by Pierce by involving more than one SWS mode and multi-stream beam.
We provide below a concise summary of the simplest case of a single-stream
electron beam coupled to a single TL representing the primary eigenmode
of the SWS.

The state of the TWT system is described by variables $Q(t,z)=\int_{t_{0}}^{t}I(t',z)dt'$
and $q(t,z)=\int_{t_{0}}^{t}i(t',z)dt'$ where $I$~and $i$ are
respectively the TL line and the electron beam currents, $t_{0}$
is the initial time. Variables $Q$ and $q$ represent the amount
of charge that has traversed the cross-section of the transmission
line and the electron beam, respectively, at point $z$, from time
$t_{0}$ to time $t$. Then, following \cite{figotin2020analytic},
we introduce the TWT-system Lagrangian $\mathcal{L}_{TB}$ as
\begin{equation}
\mathcal{L}_{TB}=\mathcal{L}_{Tb}+\mathcal{L}_{B},
\end{equation}
where the Lagrangian components $\mathcal{L}_{Tb}$ and $\mathcal{L}_{B}$
are associated with the SWS and the electron beam respectively and
are defined as follows

\begin{equation}
\mathcal{L}_{Tb}=\frac{L}{2}\left(\partial_{t}Q\right)^{2}-\frac{1}{2C}\left(\partial_{z}Q+b\partial_{z}q\right)^{2},\label{eq:Lagrangian TL}
\end{equation}
\begin{equation}
\mathcal{L}_{B}=\frac{v_{0}^{2}\partial_{z}q\partial_{t}q}{2\beta}-\frac{2\pi}{\sigma_{B}}q^{2}.\label{eq:Lagrangian Beam}
\end{equation}

Here, $\sigma_{B}$ is the cross-section of electron beam and $v_{0}$
is the electron beam stream velocity. The symbols $\partial_{t},\partial_{z}$
represents the partial derivative with respect to time $t$ and space
$z$, respectively. The parameters $L$ and $C$ are , respectively,
the distributed inductance and capacitance associated with the single
TL. The term $b$ in \ref{eq:Lagrangian TL} describes how the electron
beam couples to the TL. The representation of the coupling between
an electron beam and a SWS goes back to Ramo \cite{ramo_currents_1939}.
The debunching effects are considered by the term $-2\pi q^{2}/\sigma_{B}$
in equation \ref{eq:Lagrangian Beam}. The parameter $\beta$ is the
electron beam stream intensity and it equals $\sigma_{B}R_{sc}^{2}\omega_{p}^{2}/(4\pi)$,
where $\omega_{p},R_{sc}$ are the corresponding plasma frequency
and plasma frequency reduction factor, respectively. The Euler-Lagrange
equations associated with the Lagrangian are the following system
of second-order differential equations:

\begin{equation}
L\partial_{t}^{2}Q-\frac{1}{C}\left(\partial_{z}^{2}Q+b\partial_{z}^{2}q\right)=0,\label{eq:Euler 1st}
\end{equation}

\begin{equation}
\frac{1}{\beta}\left(\partial_{t}+v_{0}\partial_{z}\right){}^{2}q+\frac{4\pi}{\sigma_{B}}q-\frac{b}{C}\left(\partial_{z}^{2}Q+b\partial_{z}^{2}q\right)=0.\label{eq:Euler 2nd}
\end{equation}

Since the beam parameters are assumed constant in space, we can make
use of the dispersion relation to study the eigenmodes. With that
in mind, we consider solutions of the form $q(z,t),\:Q(z,t)\propto e^{j(\omega t-kz)}$.
In the case of spatially uniform (homogeneous) TL, the Fourier transform,
in time $t$ and space variable $z$, of equations \ref{eq:Euler 1st},
\ref{eq:Euler 2nd} yields
\begin{equation}
\left(\frac{k^{2}}{C}-\omega^{2}L\right)\hat{Q}+\frac{k^{2}}{C}b\hat{q}=0,\label{eq: FT Linear eqn 1st}
\end{equation}

\begin{equation}
\left[\frac{4\pi}{\sigma_{B}}-\frac{1}{\beta}\left(\omega-v_{0}k\right)^{2}\right]\hat{q}+\frac{k^{2}}{C}\left(b^{2}\hat{q}+b\hat{Q}\right)=0,\label{eq: FT Linear eqn 2nd}
\end{equation}
where $\omega$ and $k$ are respectively the frequency and the wavenumber,
and $\hat{Q}=\hat{Q}(\omega,k)$ and $\hat{q}=\hat{q}(\omega,k)$
are the Fourier transforms of the system variables $Q(t,z)$ and $q(t,z)$,
respectively. The above system of linear equations \ref{eq: FT Linear eqn 1st},
\ref{eq: FT Linear eqn 2nd} is of special interest to us as for it
encodes important information on the TWT system including its dispersion
relations and the structure of the eigenmodes. For every fixed $\omega$,
the linear system \ref{eq: FT Linear eqn 1st}, \ref{eq: FT Linear eqn 2nd}
are viewed as a kind of eigenvalue problem, which is not the standard
eigenvalue problem, where the $k$ is an eigenvalue and the pair $\hat{Q},$$\hat{q}$
forms an eigenvector, see details in \cite{figotin2020analytic}.
Taking into account the significant role played by the velocities
in electron flow interactions, we recast equations \ref{eq: FT Linear eqn 1st},
\ref{eq: FT Linear eqn 2nd} by substituting there $k=\omega/u$ where
$u$ is the phase velocity of a mode in the interactive system. We
then solve the system of equations \ref{eq: FT Linear eqn 1st}, \ref{eq: FT Linear eqn 2nd}
assuming non-trivial (non-zero) solutions, and after elementary transformations
we arrive at the following dimensionless form of the dispersion relation
\begin{equation}
D_{s}(u,\omega)=\frac{\left(v_{0}-u\right)^{2}}{u^{2}}+\frac{\gamma}{w^{2}-u^{2}}-\frac{R_{sc}^{2}\omega_{p}^{2}}{\omega^{2}}=0,\label{eq:dispersion eqn}
\end{equation}
where $\gamma=\beta b^{2}/C$ is the system coupling parameter \cite{figotin2020analytic}
with unit of {[}velocity{]}$^{2}$, and $w=1/\sqrt{LC}$ is the cold
SWS mode phase velocity (i.e., TL model). The Euler-Lagrange equations
\ref{eq:Euler 1st}, \ref{eq:Euler 2nd} and the system of second-order
differential equations \ref{eq: FT Linear eqn 1st}, \ref{eq: FT Linear eqn 2nd}
are written in centimeter\textendash gram\textendash second (CGS)
system, whereas the dispersion relation is dimensionless, hence in
the following we will use SI units for convenience. In SI units, the
parameters $R_{sc}$ and $\omega_{p}$ are given in \cite{doi:10.1063/5.0051462}.
The same dispersion relation was also obtained by using a method based
on the Pierce model including the space charge effect. The translation
between the Lagrangian model parameters used in this framework and
the parameters used in Pierce model is listed in the Supplementary
Material of \cite{doi:10.1063/5.0051462}.

The dispersion equation \ref{eq:dispersion eqn} has three main parameters:
(i) the electron stream velocity $v_{0}$ which is chosen to be equal
the electron beam particles' initial velocity; (ii) the SWS TL modal
phase velocity $w$, and (iii) the system coupling parameter $\gamma$.
Recovering the correct values of these parameters as functions of
the angular frequency $\omega$ is not straight forward, and the purpose
of this paper is to develop an approach that uses values of these
parameters without any recourse to ``curve fitting'' approximations
but rather based on simple cold SWS full-wave simulations and a single
(at one frequency) three dimensional PIC simulation as explained next.

\subsection*{Analytical model adjustment}

In this subsection, we present a way of improving the agreement between
the analytical model previously discussed and real device data by
introducing a phenomenological adjustment to the analytical model.
The main physics-based modification is the introduction of frequency
dependence of parameters $w$ and $\gamma$ \ref{eq:dispersion eqn}
that were assumed to be constant in the analytic model. In other words,
we adjust the analytical model by assuming that $w=w(\omega)$ and
$\gamma=\gamma(\omega)$ in \ref{eq:dispersion eqn}, and that yields
\begin{equation}
D_{s}(u,\omega)=\frac{\left(v_{0}-u\right)^{2}}{u^{2}}+\frac{\gamma(\omega)}{w^{2}(\omega)-u^{2}}-\frac{R_{sc}^{2}\omega_{p}^{2}}{\omega^{2}}=0.\label{eq:adjust disperion equation-1}
\end{equation}
The frequency dependence of analytical model parameters ($w$ and
$\gamma$) is invoked as an ad-hock to the final characteristics equation
in \ref{eq:dispersion eqn}. In the following sections, we present
an approach to recover the frequency dependent parameters by exploring
solution sets of the dispersion equation \ref{eq:dispersion eqn}
at one or more frequencies and comparing that to the hot complex-valued
eigenmode observations obtained based on PIC simulations.

\section{TWT hot eigenmodes based on PIC simulations \label{sec:Hot-structure-eigenmodes}}

\begin{figure*}[t]
\begin{centering}
\centering\includegraphics[width=0.7\paperwidth]{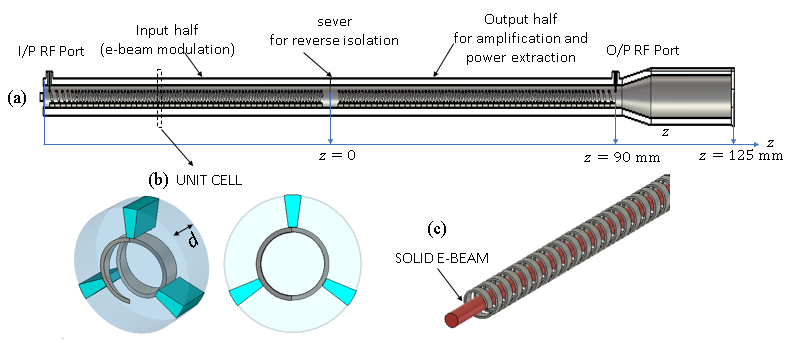}
\par\end{centering}
\caption{\label{fig:Helical-TWT,-cold sim}(a) Schematic of a helix-based TWT
amplifier. (b) (left) Perspective view and (right) front view of a
unit cell of the TWT shown in (a) where the tape helix SWS in the
circular waveguide supported by three equally spaced dielectric rods.
(c) A solid electron beam flows along the axis of the helical conductor.}
\end{figure*}

Our goal here is to develop an approach of recovering the TWT hot
eigenmode information by a thoughtful selection of regimes of TWT
operation. The selected regime carries TWT eigenmode information in
the most pronounced and undisturbed form. The approach features the
estimation of the hot eigenmode information, such as $(\omega,k),$where
$\omega$ and $k$ are the angular frequency and the associated complex-valued
wave number, respectively, based on one PIC simulation where we extract
particles information at one frequency. In general, there is no straightforward
way to extract an information about the eigenmodes from commonly performed
PIC simulation.

\subsection{Eigenmode-like operation regime}

The eigenmode-like regime of operation is determined by a set of conditions
that facilitate the manifestation of the eigenmode frequency and wavenumber
dispersion through analyzing observable parameters in PIC simulations.
Those conditions include, first, limiting our setup to the region
of frequencies where the cold SWS has a single dominant mode. Consequently,
the 4-wave theory based on Lagrangian formalism in Sec. \ref{sec:Analytical-model}
is expected to provide an accurate account for the interaction between
the electron beam and the EM wave in the SWS. Second, we would like
to suppress backward waves by all means available including introducing
a sever. Third, we choose TWT regimes of operation for which nonlinear
effects are minimal (i.e., negligible). Suppose now that the above
conditions are implemented. Then we carry out PIC simulation assuming
that the complex time representation of a chosen observable $s$ (for
instance, electric or magnetic field) after reaching the steady state
can be represented as follows:

\begin{equation}
s(t,z)=\sum_{i=1}^{4}a_{i}\exp\left\{ j\left(\omega t-k_{i}z\right)\right\} ,\label{eq:physical quatity}
\end{equation}
where $\exp\left\{ j\left(\omega t-k_{i}z\right)\right\} $ represents
the TWT hot eigenmodes, $\omega$ and $k_{i}=k_{i}(\omega)$ are the
associated angular frequency and complex-valued wavenumber, respectively.
The four modes are the resultant of the interaction between two EM
waves in the cold SWS (along $\pm z$) along with two electron beam
space-charge waves.

In a simple fully interactive system, four hot eigenmodes form the
basis for two regions of operation
\begin{enumerate}
\item Amplification region: in this region the four modes are divided into
two sets of modes: the first set consists of two exponentially growing
and decaying oscillatory modes (amplifying/attenuating, $\Im(k)\neq0$)
such that two modes wavenumbers are complex conjugate to each other
(i.e., $k_{1}=k_{2}^{*}$); the second set consists of two oscillatory
modes (unamplifying/unattenuating, $\Im\left(k\right)=0$) that vary
harmonically in time and are bounded in the entire space by a constant.
\item Non-amplified region: in this region the four modes are oscillatory
with real-valued wavenumbers (i.e. $\Im\left(k_{1:4}\right)=0$).
\end{enumerate}
\textit{Assumption 1}: In the amplification region the amplifying
eigenmode is the dominant one. Such that, for a large value of $z$
and after reaching the steady state any observable physical quantity
in \ref{eq:physical quatity} can readily be approximated as follows:

\begin{equation}
s(t,z)\approx a\exp\left\{ j\left(\omega t-k_{i}z\right)\right\} ,\label{eq:approx_domaint mode}
\end{equation}
where subscript $i$ here refers to only the mode with $\Im\left\{ k_{i}\right\} >0$,
so that the mode $\exp\left\{ j\left(\omega t-k_{i}z\right)\right\} $
is amplifying and exponentially growing in space. In view of Assumption
1, the structure hot amplified eigenmode $\exp\left\{ j\left(\omega t-k_{i}z\right)\right\} $
is detected by considering the Fourier transformation in spatial variable
$z$ of one or more of the observable physical quantities in the PIC
simulation. The PIC algorithm uses a three dimensional self-consistent
solution of Maxwell\textquoteright s equations in the time domain
in the presence of an electron beam current made of by a finite number
of emitted electrons. It also accounts for saturation effects such
as electron overtaking. In the following subsection, we show an example
of helix-based SWS and the estimation of the amplified eigenmode using
observable physical quantities.

\subsection{Estimation of the amplified eigenmode in helical TWT}

We illustrate the efficiency of the approach described in the previous
section by considering an example of a C-Band TWT amplifier shown
in Fig. \ref{fig:Helical-TWT,-cold sim}(a). A helix-based SWS typically
consists of a metallic tape-helix inside a metallic waveguide; such
SWSs have been used for decades for high power device sources and
amplifiers \cite{gilmour_principles_1994,tsimring_electron_2006,gilmour_klystrons_2011}.
Figure \ref{fig:Helical-TWT,-cold sim}(a) shows an example of helix-based
TWT optimized to operate at around 15 GHz, with a total length of
about 18 cm comprised of 160 unit cells with period $d$ = 1.04 mm.
TWT unit cells are made of a helix metallic tape with an inner radius
of 795 $\mu$m, 0.2 mm thickness and 0.51 mm width. The metallic circular
waveguide has a radius of 1.06 mm and the three equally spaced dielectric
rods support that physically hold the helix are made of BeO dielectric
with a relative dielectric constant of 6.5. The input and output radio
frequency (RF) signals of the structure are defined as input RF port
and output RF port as shown in Fig. \ref{fig:Helical-TWT,-cold sim}(a).

We show in Fig. \ref{fig:Helical-TWT,-cold sim-1} the dispersion
relations of the modes in the periodic cold SWS. The cold SWS dispersion
diagram indicates that the frequency range of the first quasi-TEM
forward mode, which is the reasonable for amplification regime of
operation, ends around 25 GHz.

The ($\omega$-$k$) cold dispersion is calculated using the eigenmode
solver implemented in CST Suite Studio by DS SIMULIA based on the
finite-element method. The eigenmode solver enforces a phase shift
across the structure period in the longitudinal direction of propagation
and solves for the real-valued eigenfrequencies. The dispersion diagram
is constructed by repeating the simulation for different phase shifts.

The cold dispersion diagram shows both forward and backward Floquet-Bloch
propagating harmonics in the fundamental Brillouin zone that is here
defined from $kd/\pi=0$ to $kd/\pi=2$. The plotted dispersion curves
represent only the propagating part of the spectrum and thus, they
have a purely real-valued wavenumber $k$. In addition, it is worth
emphasizing that the dispersion diagram obtained in Fig. \ref{fig:Helical-TWT,-cold sim}(b)
is for the cold SWS case (electron beam is absent). However, this
cold dispersion is still helpful as a 0th-order approximation to establish
synchronization in the fully interactive system (SWS and electron
beam). This approximate picture of the interactive system is realized
by plotting the \textquotedblleft electron beam line\textquotedblright{}
together with the cold dispersion diagram as shown in Fig. \ref{fig:Helical-TWT,-cold sim-1}
in the blue-dashed line. As shown in Fig. \ref{fig:Helical-TWT,-cold sim-1},
the electron beam line intersects with the TEM eigenmode in red around
12 GHz.

\begin{figure}[htbp]
\begin{centering}
\centering\includegraphics[width=0.7\columnwidth]{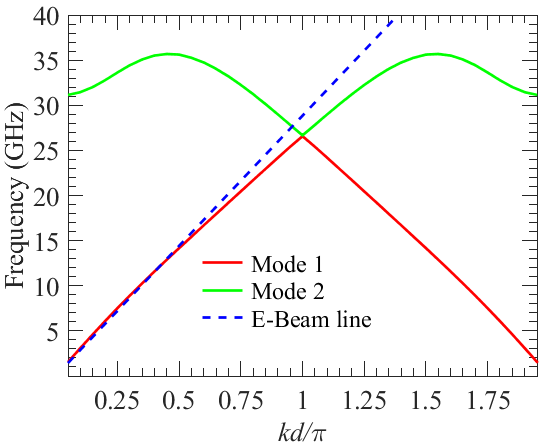}
\par\end{centering}
\caption{\label{fig:Helical-TWT,-cold sim-1} Dispersion diagram of the lowest
order Floquet-Bloch eigenmodes of the cold SWS in Fig. \ref{fig:Helical-TWT,-cold sim}.
An electron beam (e-beam) line with particles' initial velocity about
0.2 times the speed of light to have a synchronization around the
12 GHz (blue dashed line) is plotted showing that the beam line intersects
with the cold dispersion in the forward region.}
\end{figure}

The hot simulation setup is carried out by CST Studio Suite (PIC solver).
The PIC algorithm uses a 3-D self-consistent solution to Maxwell\textquoteright s
equation in the time domain in the presence of an electron beam current
made by a finite number of emitted charged particles. It also accounts
for saturation effects such as electron overtaking which is responsible
for reaching a steady state in an oscillator.

The hot simulation is performed taking into account the eigenmode-like
regime considerations by defining particles that are emitted based
on the direct current (DC) emission model of the PIC Solver. Such
TWT amplifier uses a solid linear electron beam with a radius of 560
$\mu$m. The particles' initial velocity is 0.2 times the speed of
light $c$ (i.e., $v_{0}$ = 0.2$c$) to have a synchronization around
12 GHz in the region where the cold SWS unit cell can be modeled by
one cold eigenmode (in each direction) as shown in the cold dispersion
and to have a good matching to minimize any rise of the backward (i.e.,
reflected) modes due to mismatch at the RF ports. The value of the
emitted current was set to 10 mA that is less than the threshold current
for oscillation, to minimize nonlinear effects. A static axial magnetic
field $B_{z}$= 0.64 T is used to ensure a good beam confinement of
the solid electron beam traveling in the axial direction. The total
number of charged particles used to model the electron beam in the
PIC simulation is about 8,901,500 while the whole space in the SWS
structure is modeled using 6,640,704 mesh cells.

\begin{figure}[t]
\begin{centering}
\centering\includegraphics[width=0.65\columnwidth]{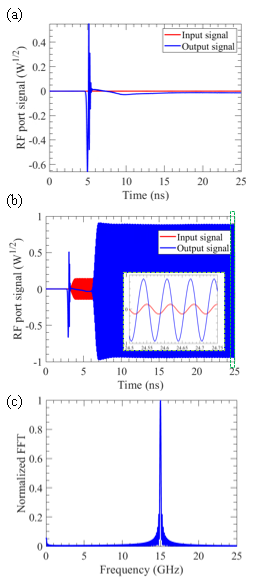}
\par\end{centering}
\caption{\label{fig:Hot_simulation_input_output_ports} (a) Simulated output
RF port signal, for the no RF excitation case, shows no signal after
some transient time \textasciitilde{} 10 nS, confirms the stability
of the structure, and indicates the time required to reach steady
state. (b) Simulated output RF port signal in green shows an amplification
occurs to the input RF signal of 10 dBm input and 15 GHz in red, (c)
frequency spectrum of the output RF signal shows that the output signal
has the same frequency of the signal used to excite the input RF port
but with a higher amplitude indicates the amplification operation.}
\end{figure}

A PIC simulation is performed to verify the stability of the structure
and its immunity to oscillation by running the CST PIC for no RF excitation
at the input RF port (red trace), and the signal observed at the output
port (blue trace) is shown in Fig. \ref{fig:Hot_simulation_input_output_ports}(a).
It is clear that in the case where there is no RF excitation and since
the hot setup configuration, that ensures the stability of the hot
structure, the output vanishes after passing the transient time.

Figure \ref{fig:Hot_simulation_input_output_ports}(b) shows the output
(in blue) and input (in red) RF signals, indicating the amplification
level for a single-tune sinusoidal excitation signal of frequency
15 GHz. The frequency spectrum of the output RF signal shown in Fig.
\ref{fig:Hot_simulation_input_output_ports}(c) is obtained by applying
Fourier transform to the output RF signal in the time window from
10 nsec to 25 nsec. By utilizing the eigenmode-like regime of operation
in the hot setup, one can assume that any observable physical quantity
after reaching the steady state resembles the hot amplified eigenmode
as in \ref{eq:approx_domaint mode}.

To estimate the amplified hot eigenmode, we analyze the observable
physical quantities obtained in the CST PIC simulations. The PIC solver
simulates the complex interaction between the EM wave and electron
beam using a large number of charged particles and it follows their
trajectories in self-consistent electromagnetic fields computed on
a fixed mesh. The physical quantities that represent the EM mode in
the interacting SWS are electric and magnetic fields. Also, one can
observe physical quantities related to the electron beam such as beam
electrons' energy, momentum, and electron beam charge density.

\begin{figure}[tbh]
\begin{centering}
\centering\includegraphics[width=0.65\columnwidth]{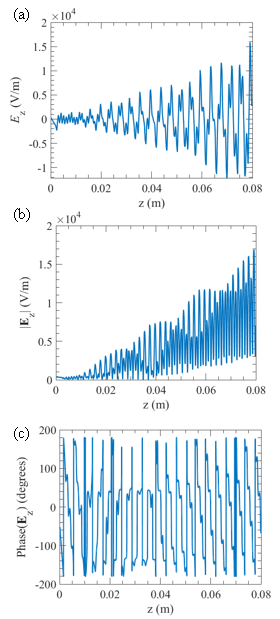}
\par\end{centering}
\caption{\label{fig:Electric-field-along}(a) A snapshot in time at $t=10$
nS of the transient $\mathbf{E}_{z}$-field just outside the helix
along the SWS after reaching steady state time. (b, c) Phasor $\mathbf{E}_{z}$-field
in terms of amplitude and phase along the SWS.}
\end{figure}

Figure \ref{fig:Electric-field-along}(a) shows a snapshot of the
$z$-component of the transient electric field along the SWS after
the steady state is reached (on the gap between the SWS and circular
waveguide wall). The phasor-domain representation of the $E_{z}(z,t)$
versus $z$ along the SWS is calculated as
\begin{equation}
\mathbf{E}_{z}(z)=\frac{1}{T}\int_{t=t_{0}}^{t=t_{0}+T}E_{z}(z,t)e^{-j\omega t}dt,\label{eq:Phasor calculations}
\end{equation}
where $t_{0}$ is any time instant after steady state is observed,
and $T=2\pi/\omega$. The phasor $\mathbf{E}_{z}(z)$ is depicted
in Fig. \ref{fig:Electric-field-along}(b, c) in terms of magnitude
and phase, respectively.

\begin{figure}[H]
\begin{centering}
\centering\includegraphics[width=0.65\columnwidth]{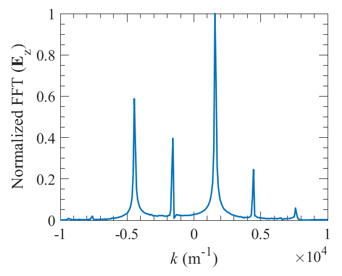}
\par\end{centering}
\caption{\label{fig:Electric-field-along-1} Spatial frequency components of
longitudinal electric field associated to the hot structure obtained
by performing FFT of phasor $\mathbf{E}_{z}(z)$ shown in Fig. \ref{fig:Electric-field-along}
in the space window from $z=0$ to $z=0.08$ m.}
\end{figure}

It is apparent that there is more than one spatial frequency component
in the field data shown in Fig. \ref{fig:Electric-field-along}. Those
spatial frequency components are obtained by taking the Fourier transform
of the $\mathbf{E}_{z}(z)$. We report the normalized spatial spectrum
in Fig. \ref{fig:Electric-field-along-1} showing different spatial
frequency components where the fundamental one that carries most of
the energy is $k=1580\,\textrm{m}^{-1}$. Also, it is worth noting
that, the plotted normalized spatial spectrum in Fig. \ref{fig:Electric-field-along-1}
invokes other modes from PIC simulations that are not part of the
analytical model eigenmodes solution. Those other modes are related
to the Floquet-Bloch modes associated with the periodic SWS such as
$k_{n}=k_{0}+2\pi n/d$ where $n$ is an integer number representing
the order of harmonics. For the particular illustrative example shown
here, we report that $k_{0}=1580\,\textrm{m}^{-1}$ and $2\pi/d=6041.5\,\textrm{m}^{-1}$
accordingly the corresponding $-1$ Floquet-Bloch harmonic $k_{-1}=-4461.5\,\textrm{m}^{-1}$,
similarly for the negative fundamental $k_{0}=-1580\,\textrm{m}^{-1}$
the corresponding $+1$ Floquet-Bloch harmonic is $k_{1}=4461.5\,\textrm{m}^{-1}.$

Similarly, we apply the same procedure to a physical quantity related
to the electron beam such as electrons' energy. In reality, the energy
of the electrons in the beam does not have a single value at each
$z$ point. The energy usually depends on each electron transverse
location, hence it is convenient to deal with the average of all electrons'
energies at each transverse $z$-dependent cross section such that
at each point $z$ the electrons' energy is represented by a single
value representing their average, as was done for the beam electrons'
speed in \cite[Eqn. 6]{mealy2020traveling}. 

\begin{figure}[t]
\begin{centering}
\centering\includegraphics[width=0.65\columnwidth]{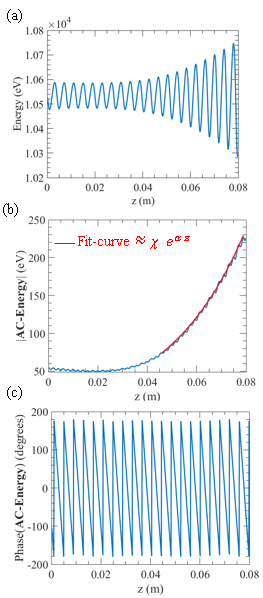}
\par\end{centering}
\caption{\label{fig:(a)-A-snapshot energy}(a) A snapshot in time at $t$ =
10 nS of the average beam electrons' energy along the SWS after reaching
steady state time. (b, c) Complex-valued AC part of the phasor beam
electrons' energy in terms of amplitude and phase. Red line in (b)
represents the exponential curve $\chi e^{\alpha z}$ with $\alpha=31.6\,\textrm{m}^{-1}$
and $\chi=19.7\,\textrm{eV}$ which fits well to the magnitude growth
of the AC part of beam electrons' energy.}
\end{figure}

Figure \ref{fig:(a)-A-snapshot energy}(a) shows a snapshot of the
average kinetic energy of the beam's electrons along the SWS after
steady state is reached. The phasor-domain representation of small-signal
kinetic energy (i.e., by subtracting the time-averaged kinetic energy)
of the electron beam ``AC-Energy'' is calculated similarly to the
$E_{z}$ in \ref{eq:Phasor calculations}. The phasor small-signal
kinetic energy is depicted in Fig. \ref{fig:(a)-A-snapshot energy}(b,
c) in terms of magnitude and phase, respectively.

\begin{figure}[htbp]
\begin{centering}
\centering\includegraphics[width=0.65\columnwidth]{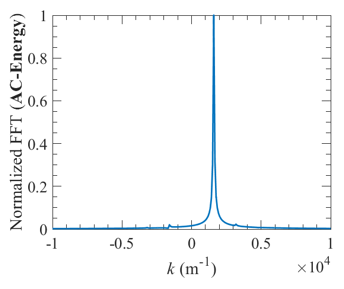}
\par\end{centering}
\caption{\label{fig:(a)-A-snapshot energy-1} Spatial frequency components
of electron beam energy associated to the TWT obtained by performing
FFT of phasor \textbf{AC-Energy} shown in Fig. \ref{fig:(a)-A-snapshot energy}
in the space window from $z=0$ to $z=0.08$ m.}
\end{figure}

It is apparent that there is only one spatial frequency component
in the electron beam data shown in Fig. \ref{fig:(a)-A-snapshot energy},
which is contrary to the electric field spectrum that involves Floquet-Bloch
modes due to the periodicity nature of the SWS. We report the normalized
spatial spectrum of phasor small-signal kinetic energy in Fig. \ref{fig:(a)-A-snapshot energy-1}
where its spatial frequency is $k=1595\,\textrm{m}^{-1}$. This spatial
frequency component is directly related to the eigenmode that is harvesting
energy from the electron beam.

By analyzing the spatial frequency spectrum of such physical observables
at a certain frequency one can estimate the dominant amplified eigenmode
complex-valued wavenumber and the corresponding complex-conjugate
attenuated eigenmode. For the helical TWT understudy and from the
spatial frequency spectrum in Fig. \ref{fig:(a)-A-snapshot energy-1}
and the amplitude growth rate reported in Fig. \ref{fig:(a)-A-snapshot energy}(b)
the estimated eigenmodes are $(\omega,k)\sim(15\textrm{GHz},1595\pm j31.6\textrm{m}^{-1}$).

\section{Connection between the analytical model and actual electron beam
device\label{sec:Connection-between-the}}

TWT observables, like EM fields and beam electrons' energy, are used
to determine the analytical model parameters ($\gamma$, $w$) through
full-wave simulations either for the model with constants parameters
or the adjusted model with dispersive model parameters. As to their
qualification to be TWT observables, we notice that by their very
definition these observables are some of the parameters associated
to the amplified eigenmode in the fully interactive system, since
the amplified eigenmode is the dominating the others for \textit{z}-locations
away from the input port. The wavenumber-frequency dispersion describing
the eigenmodes in the TWT is determined by running multiple PIC simulations
at different frequencies and then determining hot amplified eigenmode
($\omega,$~$k_{PIC}$) at each. Those eigenmodes can be observed
through a series of PIC simulations each where the structure is excited
by a single tone RF signal. Notice also that TWT observables give
knowledge about the complex-valued wavenumbers $k$ for certain frequency
$\omega$ at which the TWT dispersion relations as well the TWT characteristic
function $D_{s}(\omega,\,k)$ solutions attain instabilities (i.e.
convective instability referring to growing and decaying waves with
space) as discussed in Sec. \ref{sec:Analytical-model}.
\begin{figure*}[tbh]
\begin{centering}
\includegraphics[width=0.7\paperwidth]{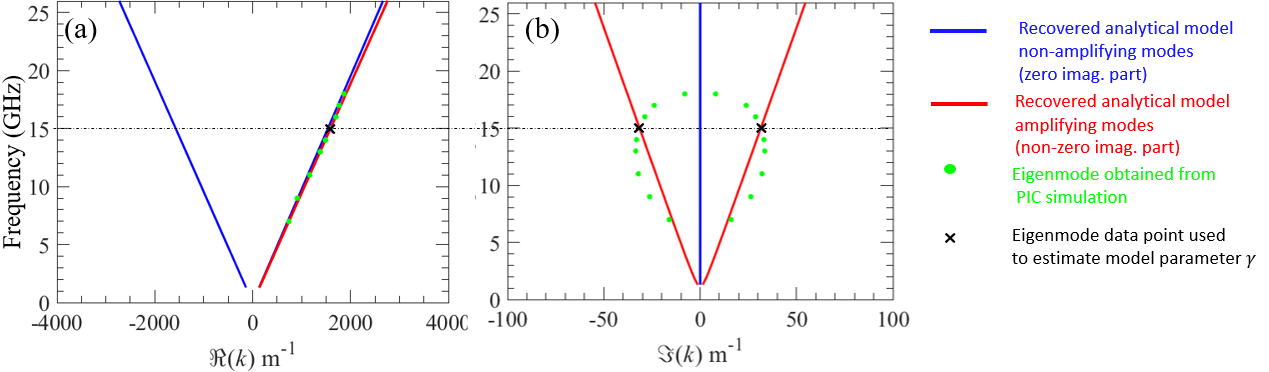}
\par\end{centering}
\caption{\label{fig:Recoverd_Dispersion}Recovered dispersion diagram, assuming
the model parameters are frequency independent. Complex-valued wavenumbers
of the eigenmodes in the interactive electron beam-EM mode in the
helical SWS in Fig. \ref{fig:Helical-TWT,-cold sim}(a). (a) Real
part of the recovered dispersion by estimating analytical model parameters
using PIC simulations data, a reasonable agreement is observed in
the real part. (b) Imaginary part of the dispersion shows a discrepancy
between the recovered model and PIC simulations data. The use of a
simple analytical model that has only non-dispersive parameters leads
to the imaginary parts of the amplified/attenuated modes (red straight
lines in (b)) which are not in consistent with PIC simulations data.}
\end{figure*}

To facilitate the computation of analytical model parameter $\gamma$
we would like to optimize $\gamma$ to best fit growing solutions
of the dispersion relation \ref{eq:dispersion eqn} $(\omega,\,k_{D_{s}}=\omega/u)$
to the estimated amplified eigenmode through PIC simulations $(\omega,\,k_{PIC})$.
This optimization process is applied only at the ``estimation frequency''.
The estimation frequency can be any frequency in the vicinity of the
synchronization frequency, within the amplification region. A common
way to deal with this kind of problem is to look for analytical parameter
$\gamma$ that minimizes error between the analytical model $k_{D_{s}}$
and PIC $k_{PIC}$ data for the estimated frequency $\omega$, by
defining the following error function: 
\begin{equation}
\mathscr{E}=c_{1}\left|\Re\left(k_{D_{s}}-k_{PIC}\right)\right|{}^{2}+\left|\Im\left(k_{D_{s}}-k_{PIC}\right)\right|{}^{2},\label{eq:error}
\end{equation}
where $c_{1}$ is a weighting coefficient to equate the importance
of the real and imaginary parts of the wavenumbers since there are
orders of magnitude difference between the real and imaginary parts
of $k$. In particular, given the aforementioned estimated eigenmode
$(\omega,\,k_{PIC})\sim(15\textrm{GHz},\,1595+j31.6\textrm{m}^{-1}$),
the constant $c_{1}$ was set to be equal to the ratio between the
imaginary and real parts of $k$, i.e., $c_{1}=$31.6/1595. Then the
expression of the error function is minimized numerically by optimizing
the analytical parameter $\gamma$.

For the particular illustrative example shown here, the parameters
are as follows: $v_{0}=w=0.2c$ are the parameters (electrons velocity
and cold EM mode phase velocity) that are established initially for
the synchronized regime of operation, whereas the optimized analytical
parameter is $\gamma=8.651\times10^{10}\,\textrm{\ensuremath{\textrm{m}^{2}}}/\textrm{s}^{2},$
where $\gamma$ is optimized based on the PIC data at 15 GHz (e.g.,
a frequency at or near the synchronization point) by minimizing the
error function. In Fig. \ref{fig:Recoverd_Dispersion}, blue and red
solid lines show the recovered dispersion for the optimized analytical
parameters. For the sake of comparison and validation of the proposed
model, we also perform PIC simulations at different frequencies and
follow to the approach described in Sec. \ref{sec:Hot-structure-eigenmodes}
to estimate the hot amplified eigenmode for each frequency. The PIC-based
hot amplified eigenmodes are depicted in Fig. \ref{fig:Recoverd_Dispersion}
in green circle dots. A reasonable agreement is observed between the
dispersion obtained by the proposed model and actual data from PIC
simulations, for the real part of the dispersion in Fig. \ref{fig:Recoverd_Dispersion}(a).
Whereas, the imaginary part of the dispersion shown in Fig. \ref{fig:Recoverd_Dispersion}(b)
shows a discrepancy except at 15 GHz ``black cross symbol'' which
is the CST PIC data point used to optimize $\gamma$. The discrepancy
at the other frequency points can be explained as the result of using
a simple analytical model that has only non-dispersive parameters
(i.e., $\gamma,v_{0},\textrm{and }w$ defined as frequency-independent
which is not practically true). The model is easily improved as described
in the following subsection, where the frequency variation of two
fundamental cold SWS parameters is accounted for.

\subsection*{Adjusted analytical model}

\begin{figure}[tbh]
\begin{centering}
\includegraphics[width=0.85\columnwidth]{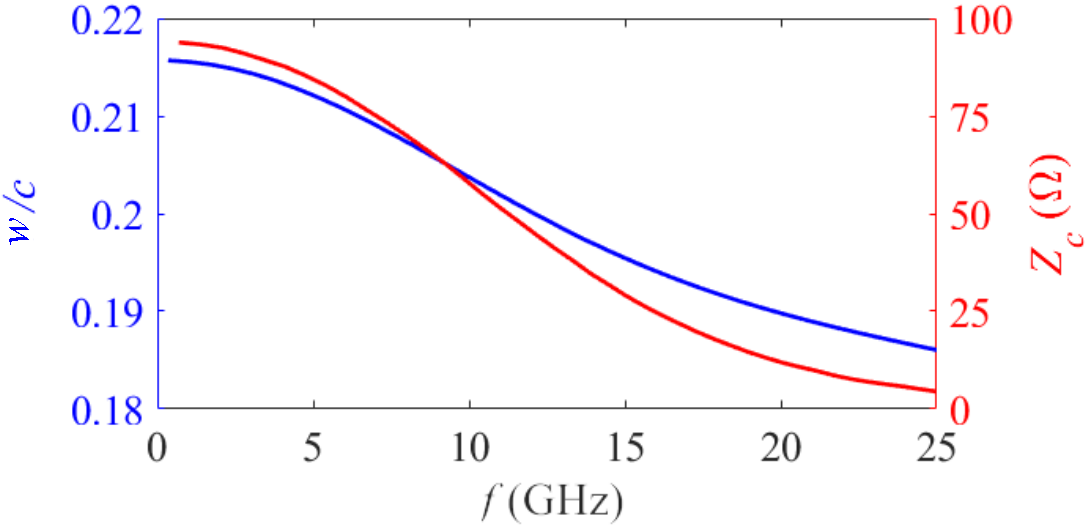}
\par\end{centering}
\caption{\label{fig:Phase velcoity} (Blue) Normalized phase velocity and (Red)
characteristic impedance of the first forward mode of the SWS shown
in Fig. \ref{fig:Helical-TWT,-cold sim}(a) obtained from cold full-wave
simulations in the absence of the electron beam, using the finite
element method-based eigenmode solver. }
\end{figure}

In this subsection, we show an improved matching between the analytical
model and PIC results attributed to the adjusted analytical model.
The adjustment is the replacement of constants $w$ and $\gamma$
with the corresponding frequency dependent functions $w(\omega)$
and $\gamma(\omega)$. The phase velocity frequency dependence $w(\omega)$
is obtained from the cold full-wave simulation by analyzing wave propagation
in the SWS in the absence of the electron beam to get the $w(\omega)$
to be used in our analytical model recovery method. The frequency
dependence of $\gamma$ is also obtained from the cold simulation
and relation $\gamma(\omega)=B/C(\omega)$ where $C(\omega)$ is the
frequency-dependent equivalent per-unit-length capacitance of the
TL model of the SWS as discussed in \cite{doi:10.1063/5.0051462},
and $B$ is an adjustment constant which is evaluated by minimizing
the error function in (\ref{eq:error}). We simulated the helix-based
cold SWS by using the finite element-based eigenmode solver implemented
in CST Suite Studio and extracted the (i) cold-circuit EM phase velocity
$w(\omega)$ normalized to the speed of light $c$ as shown in Fig.
\ref{fig:Phase velcoity}, and (ii) the characteristic wave impedance
$Z_{c}(\omega)$ of the mode under interest (i.e., mode 1 ``red trace''
in the cold dispersion shown in Fig. \ref{fig:Helical-TWT,-cold sim-1}).
By using $w(\omega)$ and $Z_{c}(\omega)$, one obtains the equivalent
frequency-dependent distributed inductance $L(\omega)=Z_{c}(\omega)/w(\omega)$
and capacitance $C(\omega)=1/\left[Z_{c}(\omega)w(\omega)\right]$.

\begin{figure*}[tbh]
\begin{centering}
\includegraphics[width=0.7\paperwidth]{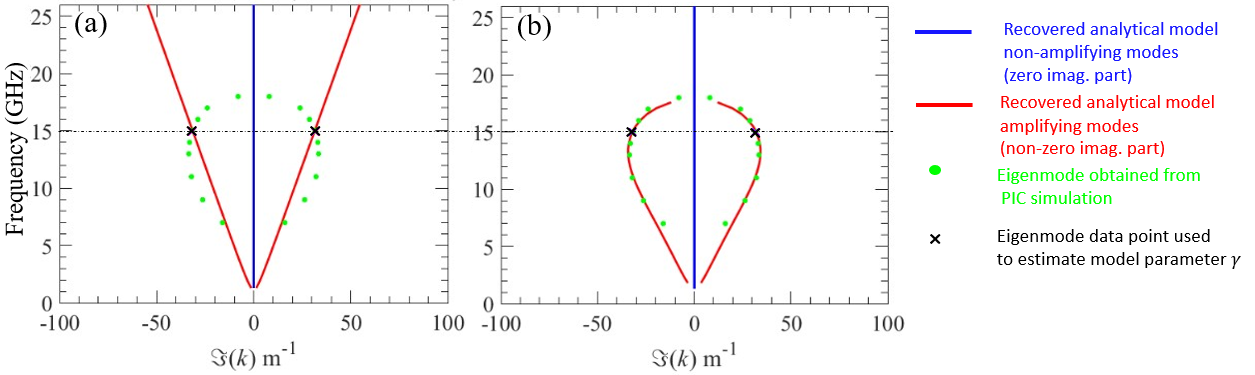}
\par\end{centering}
\caption{\label{fig:compairsion} Comparison between imaginary parts of the
wavenumbers of the hot eigenmodes for the helical SWS shown in Fig.
\ref{fig:Helical-TWT,-cold sim}(a) for two analytical models: (a)
frequency-independent parameters ($\gamma$, $w$) analytical model
in \ref{eq:dispersion eqn}, (b) frequency-dependent parameters ($\gamma(\omega)$,
$w(\omega)$) adjusted analytical model in \ref{eq:adjust disperion equation-1}.
The use of a simple analytical model that has dispersive parameters
leads to the imaginary part of the amplified/attenuated modes (red
lines in (b)) to be consistent with PIC simulations data, contrary
to what was observed when using the non-dispersive model (red lines
in (a)).}
\end{figure*}
Figure \ref{fig:After-adjust_dispersion-into} depicts the recovered
complex-valued wavenumbers dispersion of the hot modes with an electron
beam whose electrons have an initial velocity of $v_{0}=0.2c$  using
the adjusted analytical parameters as follows: the frequency dependent
phase velocity of the EM cold mode $w(\omega)$ is obtained from full-wave
cold simulations, and the optimized adjustment parameter  $B=71.28\,\textrm{F}\cdot\textrm{m}/\textrm{s}^{2}.$
Note that, $B$ is evaluated at only the ``estimation frequency''
(by minimizing the error function in \ref{eq:error} at the estimation
frequency of 15 GHz) and then the frequency dependence of $\gamma(\omega)$
is found from $\gamma(\omega)=B/C(\omega)$, where the TL distributed
capacitance $C(\omega)$ is obtained from full-wave cold simulations.
The dispersion obtained by the model is compared with the hot eigenmode
obtained directly from the PIC simulations, plotted by green circular
dots, showing a significant improvement in the agreement between the
recovered dispersion and actual data from PIC simulations when compared
to the results of the model without frequency dependent parameters.
To facilitate the comparison, 

in Figure \ref{fig:compairsion}, we plot side by side the imaginary
part of the recovered dispersion using the analytical model without
frequency dependent parameters in Fig. \ref{fig:compairsion}(a),
and the phenomenological adjusted one using frequency dispersion in
Fig. \ref{fig:compairsion}(b) to emphasize the significant improvement
of the model in matching  real data obtained from CST PIC simulations.

\section{Summary}

An adjusted analytical model for TWTs is proposed that gives accurate
predictions of the wavenumber-frequency dispersion relation and the
frequency dependent gain/amplification. The approach utilizes primary
frequency-dependent parameters of the cold SWS (modal wave velocity
and equivalent TL capacitance) recovered by using a standard procedure,
and only one PIC simulation of the TWT, at one frequency, to find
the adjustment parameter \textit{B} without curve fitting. The proposed
adjusted analytic model was tested against results obtained directly
from PIC simulation for a helical-based TWT operating in the GHz range
and showed an excellent agreement. Our extensive preliminary studies
suggest that the proposed approach can be applied to different kinds
of TWTs including those based on the serpentine SWS. The method can
also be extended to retrieve other important TWT parameters like the
plasma frequency reduction factor. 

\section{Acknowledgments}

This material is based upon work supported by the Air Force Office
of Scientific Research award number FA9550-19-1- 0103. The authors
are thankful to DS SIMULIA for providing CST Studio Suite that was
instrumental in this study.

\bibliographystyle{IEEEtran}
\bibliography{MyLibrary}

\end{document}